# A weaving process to define requirements for Cooperative Information System

Mohamed Amroune[1], Jean Michel Inglebert[2], Nacereddine Zarour[1] and Pierre Jean Charrel[2]

[1] University of Constantine, Algeria
*amroune@irit.fr, nasro-zarour@umd.edu.dz*

[2] University of Toulouse II, France
*inglebert@iut-blagnac.fr, charrel@univ-tlse2.fr*

**Abstract :**
The development of a Cooperative Information System (CIS) becomes more and more complex, new challenges arise for managing this complexity. So, the aspect paradigm is regarded as a promising software development technique which can reduce the complexity and cost of developing large software systems. This opportunity can be used to develop a CIS able to support the interconnection of organizations information systems in order to ensure a common global service and to support the tempo of change in the business world that is increasing at an exponential level.

We previously proposed an approach named AspeCiS (An Aspect-oriented Approach to Develop a Cooperative Information System) to develop a Cooperative Information System from existing Information Systems by using their artifacts such as existing requirements, and design. In this approach we have studied how to elicit CIS Requirements called Cooperative Requirements in AspeCiS. In this paper we propose a weaving process to define these requirements by reusing existing requirements and new aspectual requirements that we define to modify these requirements in order to be reused.

## 1. Introduction

Today the organizations evolve in new environments characterized by changes in customer demands, the increased competition, communications performance, etc. In order to cope with these business conditions, enterprises migrate to inter-organizational relationships [4], [5], [6] as a way to adapt to their new environment, gain competitive advantage, and, increase their efficiency. So, the enterprise cooperation is not an easy task. It requires an effective Cooperative Information System (CIS) to support this inter-enterprise cooperation.

The Software Engineering discipline has emerged, to give response to the increasing demands for software development. So, it proposed structured processes and activities to facilitate the development of software. The initial phase of Software Engineering is Requirements Engineering. We suggest improving requirements engineering in CIS through the early identification of base concerns and crosscutting concerns (that affect several modularization units). In this strategy, managing of complexity is better supported than by traditional non aspect-oriented approaches. Thus, our research aims at developing a new approach called AspeCis, which ensures the effectiveness and efficiency of business cooperation based on the Aspect concept.

In AspeCiS, when a new requirement cannot be achieved directly by an existing Information System (IS), AspeCiS composes requirements issued from other ISs in order to fulfill this requirement. The main objectives of AspeCiS are: (i) to separate existing requirements from new requirements in the CIS; (ii) to provide a high degree of functional reuse, which helps to build again the same requirements on other existing ISs. AspeCiS includes three main phases (cf. 1): (i) elicitation and analysis of CRs, (ii) models weaving (conception of CRs models), and (iii) models to code (preparation of the implementation phase).

In our previous work we have exhibited a process to elicit requirements related to the CIS to be developed. In this paper we illustrate how existing requirements can be used by a weaving process that we propose in this paper to define requirements related to the CIS to be developed.

The remainder of the paper is organized as follows: Section 2 presents an overview of AspeCiS. The AspeCiS weaving process is detailed in section 3. Section 4 draws some examples. Section 5 provides a summary of the paper and a brief overview of the continuation of this work.

## 2. AspeCiS approach: An overview

The CIS becomes more and more complex, new challenges arise for managing this complexity, for this reason general purpose methods no longer suffice. An important and particular software engineering phase we address in our research is requirements engineering (RE). The reason is that the development of a system is heavily determined by its requirement elicitation, specification and the design decision that derive from it. The later development phases will be based on the requirements elicitation and analysis.

AspeCiS propose to improve the requirements definition through the early identification of base concerns and crosscutting concerns. AspeCiS includes three main phases (cf. 1): (i) elicitation and analysis of CRs, (ii) models weaving (conception of CRs models), and (iii) models to code (preparation of the implementation phase).

Phase I: Elicitation and analysis of CRs. This phase is composed of four steps which are:
(1) the definition of CRs, (2) the refinement of CRs, (3) the formulation of CRs depending on the ERs and possibly with the definition of some aspectual requirements (ARs), (4) the selection of a set of Operators to be used to weave ERs and the ARs to define the CRs as can be seen in the figure 1.

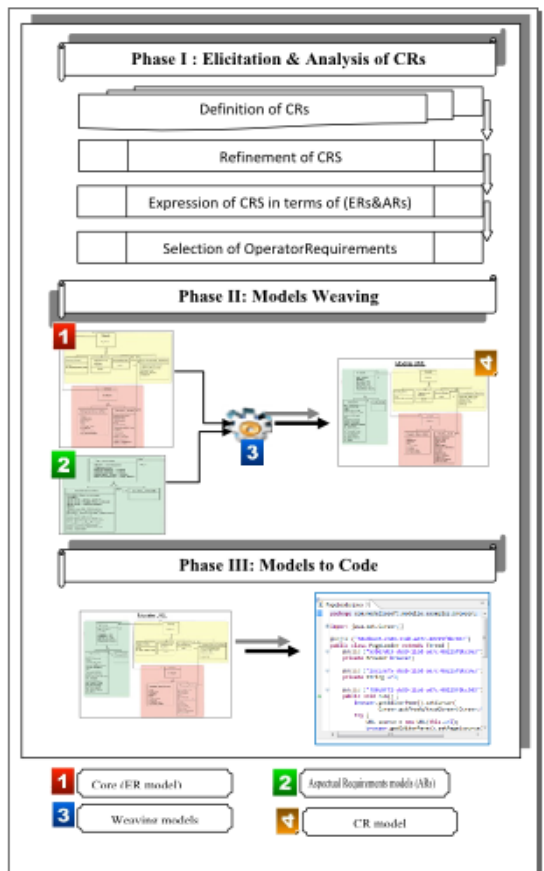

Fig. 1 Synopsis of AspeCiS

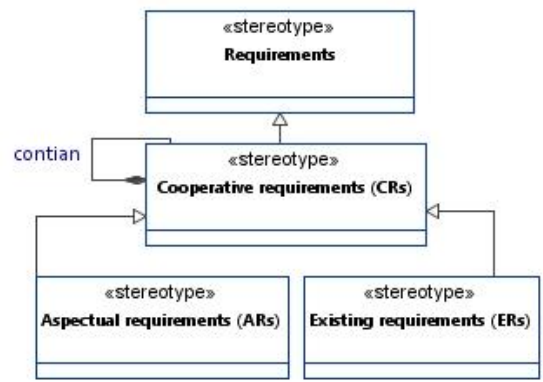

Fig. 2 Cooperative requirements metamodel

**Phase II:** Development of CRs models. The ARs should be composed with ERs models to produce CRs models. This phase includes a conflict resolution task that can appear during the requirements composition process. We proposed in [3] a conflict resolution process among aspectual requirements during the requirements engineering level: a priority value is computed for each AR, and it allows identifying a dominant AR on the basis of stakeholder priority. This process is more formal than those currently proposed, which requires a trade-off negotiation to resolve conflicts.

**Phase III:** Preparing the implementation phase. The purpose of this phase is to transform models into code templates.

2.1 The concept of Requirements in AspeCiS

Several definitions of requirement exist in the literature [7], but we adopt the following ones to differentiate between requirements in AspeCiS. So, in AspeCiS tree kind of requirements are defined.

**Existing Requirements (ERs).** They are statements of services or constraints provided by an existing system, which define how the system should react to particular inputs and how the system should behave in particular situations as shown in the figure 2.

**Aspectual Requirements (ARs).** They are concerns that cut across other existing requirements by a weaving operation in order to modify ERs to be reused to define CRs.
**Cooperative Requirements (CRs).** They are goal requirements that will be refined to relate on ERs and eventually ARs, exhibiting what parts of existing systems requirements will be reused.

In our previous work we have developed the first phase of AspeCiS that consists of a definition of cooperative requirements.

In the next section we present a weaving process that we used to develop a Cooperative Requirements (CRE) models.

## 3. A weaving process to define requirements for CIS

The weaving concept is used to support such a decoupling among models. The weaving concept is not new and the definition of model weaving considered in this paper is an extension of the generic metamodel weaving proposed by Didonet Del Fabro et al. in [8]. The general operational context of this generic metamodel weaving is depicted in Figure 3. It consists of the production of a weaving model *WM* representing the mapping between two metamodels: a left meatamodel *LeftMM* and a right metamodel *RightMM*. The *WM* model should be conform to a specific weaving metamodel *WMM*.

The generic metamodel weaving proposed by Didonet Del Fabro et al must be extended, to be used in our context. This extension (cf. figure 3), consists of the definition of a Core metamodel (*ALeftmm*), an Aspectual Requirements metamodel (*ARightmm*) and a weaving metamodel called *AWM* (for AspeCiS Weaving model) specific to our approach. So, the Core metamodel represents an ERs metamodel. It is conformed to the UML metamodel. We present in this paper the weaving model specific to AspeCiS (*AWM*).

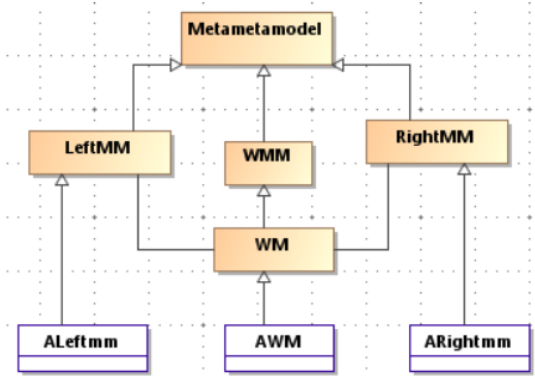

Fig. 3 Atlas model weaver AMW extension

The weaving models (*AWM*) is used to capture different kinds of links between input model elements (*ALeftmm* & *ARightmm*). The links have different semantics, depending on the application scenario. For instance (Attribute, Class) is a kind of link. It means that an attribute from Core model is added to a class from Aspectual Requirements model. The semantic of links is not in the scope of this paper.

Before presenting the *AWM* (AspeCiS metamodel), we briefly present the weaving metamodel (WM) proposed by Didonet Del Fabro et al. in [8]. This metamodel is illustrated in the Figure 4. This metamodel is composed by the following elements:

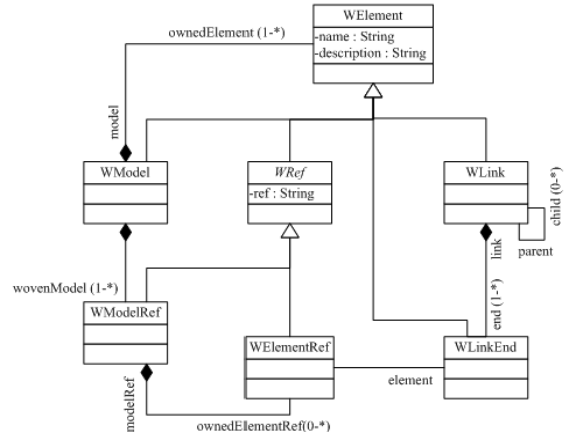

Fig. 4 Atlas Model Weaver (AMW) metamodel

- *WElement* is the base element from which all other elements inherit. It has a name and a description.
- *WModel* represents the root element that contains all model elements. It is composed by the weaving elements and the references to woven models.
- *WLink* express a link between model elements, i.e., it has a simple linking semantics. To be able to express different link types and semantics, this element is extended with different metamodels.
- *WLinkEnd* represents a linked model element.
- *WElementRef* is associated with an identification function over the related elements. The function takes as parameter the model element to be linked and returns a unique identifier for this element.
- *WModelRef* is similar to *WElementRef* element, but it references an entire model.

The *AWM* is produced and depicted in Figure 5. So, we define a model weaving which is the Weaving-Core_Aspect. It is composed of two models (Core & Aspect), as shown in the extract of KM3 following code. KM3 is a simple textual language to define metamodels [9].

```
class Weaving-Core_Aspect extends WModel {
reference Core container :
WModelRef;
reference Aspect container : WModelRef;
}
```

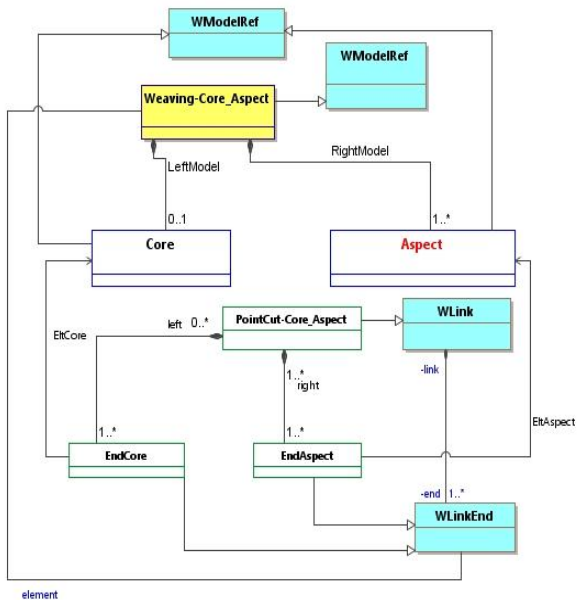

Fig. 5 AspeCiS Weaving Metamodel (AWM)

With respect to this metamodel, a Weaving-Core_Aspect consists of two models (Core model & Aspect model) related through weaving links (Pointcut-Core_Aspect). The Core model (LeftModel) is an extension of WModelRef, it represents a model of ERs. The Aspect model (RightModel) represents a model of a aspectual requirements. This model is also an extension of WModelRef.

The Weaving-Core_Aspect is also composed of the Pointcut (PointcutCoreAspect), which is an extension of WLink. The Pointcut-Core_Aspect is composed by two elements (EndAspect, EndCore), these elements are an extension of WLinkEnd. The EndCore element represents an artefact of Core model, and the EndAspect element represents an artefact of the Aspectual requirements.

## 4. The AWM Usage

In the previous sections, we have presented a weaving process to produce CRs models. It is now convenient to better illustrate the use of this process.

This example illustrates a part of the university students management system. It consists of the management of the student's subscription in the High Graduate School in Algerian universities. The Hight Graduate School composes of several universities; it aims to assure a high formation of students. After completion of courses of study, the student receives a doctor's degree.

We intend to build a CIS able to manage a cooperative project involving several universities to provide High Graduate School. Each university is supported by its existing IS. The new CIS is built on the basis of existing ISs (more details of this example are presented in [2]).

In this section, we illustrate how to weave two models (M1, M2) using AWM, in order to produce a model of CRs. At the requirements level of the existing ISs, the student subscription requirement is defined as:
**ER1**= *"Every student may have a second subscription in the same university"*. However, in the CIS, the **CR** is defined as:
**CR1**= *"Every student can have a second subscription in the same university provided that the number of hours of the second speciality does not exceed 50% of the number of hours of the first one"*.

The existing ISs allow a second subscription in the same university. So, in order to participate in the Height Graduate School, each university must respect the constraint, of the number of hours for the second subscription, imposed by the Height Graduate School's regulation. This constraint is defined in the *CR1* cited previously. Furthermore, the CIS to be developed, to support the management of this Height Graduate School, will be developed by reusing the existing ISs after some modifications.

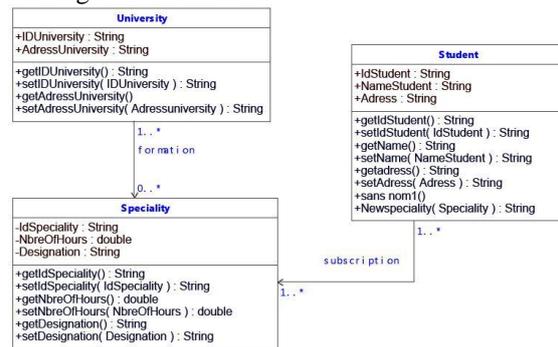

Fig. 6 A Core model

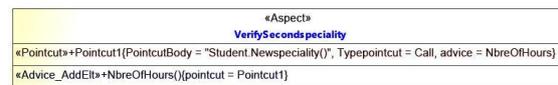

Fig 7 An Aspect model

At the model level of ISs, M1 models the *ER1* (see figure 6). It represents the core model that models the student subscription in the Height Graduate School. M1 is a class diagram conforms to UML class diagram metamodel. M1 composes of three entities which are: University, Student and Speciality. M2 (see figure 7). Represents aspet, it conforms to the AspectOperator-Requirement metamodel that we presented in [1].

In this example, the aspect model contains the advice *advice_addElt* witch role is to verify the number of hours before the call of the function *Student.NewSpeciality()*. These informations are defined in the Pointcut *Pointcut1* through the

(BodyAdvice) and the (*Typepointcut= "call"*) (see figure 6).

In this example, we use the *Weaving-Core_Aspect* to add two operations to M1, especially to the *Student* class. These operations are called before the call of the *Student.NewSubscription()* operation, in order to add a second subscription. The first operation *VerifySpecialty.NbreOfHours(IdSpecialty)* consists of the computation of the number of hours for the new subscription, the result of this operation is used by the second operation get *SecondSpecialty()* to verify the constraint imposed to authorize or not a second subscription.

## 5. Conclusion

In a previous work [2], we proposed an approach named AspeCiS to develop a Cooperative IS (CIS) from existing ISs by using their artifacts such as requirements, and design. We developed a process to elicit CRs.

In the present work, we proposed a Model driven engineering weaving process to be used to develop a cooperative e requirements models. This process is based on the use of input models which are *Core* and *Aspect* models. The *Core* models represent existing requirements, and the aspect models represent Crosscutting Requirements. These crosscutting requirements are considered as aspectual requirements and must be woven with existing requirements in order to define cooperative requirements related to the CIS to be developed. The proposed weaving metamodel is used to capture different kinds of links between model elements. These links have different semantics. We will define these semantics in our future work.

## 6. Acknowledgements

This research is partially supported by the PHC TASSILI project under the number 10MDU817. We thank the anonymous reviewers for providing valuable comments.

**Mohamed Amroune is** currently a Ph.D. Student, at the University of Toulouse II, Mirail, France and University of Tebessa, Algeria. He received his Engineer and Magister degrees in Software Engineering and Artificial Intelligence & Data Bases from the USTHB University of Algiers , Algeria, and the University of Tebessa, Algeria, in 1993 and 2007, respectively. His research interests include Information System, Requirements Engineering, Cooperation and Aspect oriented software development.

**Jean Michel inglebert** is a Doctor at the Computer Sciences Department of University of Toulouse, France. His current research activities are conducted at the IRIT laboratory, University of Toulouse.
.
**Nacereddine zarour** is a Professor at the Computer Sciences Department of University Mentouri, Constantine, Algeria. His current research activities are conducted at the LIRE laboratory, University of Constantine. He heads the project of PHC CMEP Tassili with IRIT laboratory of Toulouse 2 University. His research interests include advanced information systems, particularly cooperative information systems, architectures (based on SOA, SMA, ..), and requirements engineering.

**Pierre Jean Charrel** is a Professor at the Computer Sciences Department of University of Toulouse, France. His current research activities are conducted at the IRIT laboratory, University of Toulouse. He currently heads PHC CMEP Tassili project nr 10MDU817 with LIRE Laboratory  of Mentouri University of Constantine, Algeria. His research interests include requirements engineering and knowledge engineering, in the context of cooperative information systems.